\documentclass[preprint2]{aastex}

\usepackage{xspace}

\newcommand{\msun}{\ensuremath{M_{\sun}}\xspace}
\newcommand{\ergs}{\ensuremath{{\rm erg\,s}^{-1}}\xspace}
\newcommand{\ergscm}{\ensuremath{{\rm erg\,s^{-1}\,cm^{-2}}}\xspace}
\newcommand{\sref}{Section~\ref}

\shorttitle{A search for hyperluminous X-ray sources}
\shortauthors{Zolotukhin et al.}

\begin{document}


\title{A search for hyperluminous X-ray sources in the XMM-Newton source catalog}


\author{I. Zolotukhin\altaffilmark{1,2}, N. A. Webb\altaffilmark{3,1}, O. Godet\altaffilmark{3,1}, M. Bachetti\altaffilmark{4}, \and D. Barret\altaffilmark{3,1}}
\affil{1 CNRS; IRAP; 9 av. Colonel Roche, BP 44346, F-31028 Toulouse cedex 4, France}\email{ivan.zolotukhin@irap.omp.eu}
\affil{2 Sternberg Astronomical Institute, Moscow State University, Universitetskij pr., 13, 119992, Moscow, Russia}
\affil{3 Universit\'e de Toulouse; UPS-OMP; IRAP;  Toulouse, France}
\affil{4 INAF/Osservatorio Astronomico di Cagliari, via della Scienza 5, I-09047 Selargius (CA), Italy}


\begin{abstract}
We present a new method to identify luminous off-nuclear X-ray sources in the outskirts of galaxies from large public redshift surveys, distinguishing them from foreground and background interlopers.
Using the 3XMM-DR5 catalog of X-ray sources and the SDSS DR12 spectroscopic sample of galaxies, with the help of this off-nuclear cross-matching technique, we selected 98 sources with inferred X-ray luminosities in the range $10^{41} < L_{\rm X} < 10^{44}$\,\ergs, compatible with hyperluminous X-ray objects (HLX).
To validate the method, we verify that it allowed us to recover known HLX candidates such as ESO 243$-$49~HLX$-$1 and M82~X$-$1.
From a statistical study, we conservatively estimate that up to $71 \pm 11$ of these sources may be fore- or background sources, statistically leaving at least 16 that are likely to be HLXs, thus providing support for the existence of the HLX population.
We identify two good HLX candidates and using other publicly available datasets, in particular the VLA FIRST in radio, UKIDSS in the near-infrared, {\it GALEX} in the ultra-violet and CFHT Megacam archive in the optical, we present evidence that these objects are unlikely to be foreground or background X-ray objects of conventional types, e.g. active galactic nuclei, BL Lac objects, Galactic X-ray binaries or nearby stars.
However, additional dedicated X-ray and optical observations are needed to confirm their association with the assumed host galaxies and thus secure their HLX classification.
\end{abstract}



\section{Introduction}

Intermediate mass black holes (IMBHs) are a class of black holes (BHs) with masses greater than standard stellar mass black holes (StMBHs),
but smaller than the supermassive black holes (SMBHs) that reside in the centers of galaxies, with attributed mass of $\sim 10^6-10^9\msun$.
The masses of StMBHs are expected theoretically to span a range between roughly $\sim 5$ and $50\msun$ \citep{fryer01}.
These StMBHs are the end points of stellar evolution for sufficiently massive stars formed out of metal enriched gas.
At low metallicities, much more massive stars and black holes may be formed \citep[see, e.g.,][and references therein]{belczynski10}.
Indeed it has been suggested that zero metallicity stars can form with masses of $\sim 10^2-10^3\msun$, and produce primordial IMBH seeds directly as a result of stellar evolution \citep{heger03}.
Following processes of growth and assembly of SMBHs from these primordial seeds in a hierarchical structure formation scenario inevitably results in a population of IMBHs wandering in galaxy halos at the present epoch, with an occurrence rate of approximately 100 IMBHs per Milky Way-sized halo \citep[see e.g.][and references therein]{volonteri05}. 

Ultraluminous X-ray sources (ULXs), non-nuclear X-ray sources accreting above the Eddington limit for a StMBH ($\sim 10^{39}\,\ergs$),
have long been proposed as candidate IMBHs \citep[see][for a review]{feng11}. 
Recent studies \citep[see, e.g.,][and references therein]{sutton13,bachetti13,walton13}
have shown that the emission of most ULXs with luminosities up to $\sim 10^{41}\,\ergs$ may be explained by super-Eddington accretion onto quite massive StMBHs ($\lesssim 100\msun$).
So far only one accreting IMBH candidate, ESO 243$-$49 HLX$-$1, has been reliably identified \citep{farrell09}. 
Discovered serendipitously as a bright and variable X-ray source with maximum unabsorbed X-ray luminosity in the 0.2--10~keV band $L_{\rm X} \simeq 10^{42}\,\ergs$, it resides in an edge-on spiral galaxy some 95~Mpc away.
It is believed to be an accreting $M \sim 10^4 - 10^5 \msun$ IMBH \citep{servillat11,davis11,godet12,webb12} with a stellar companion on a 1~yr eccentric orbit (\citet{lasota11}, but see \citet{godet14}),
embedded in a young stellar system, e.g. open cluster \citep{farrell12}. 

Currently, many research teams are pursuing the search of other candidate IMBH sources similar to ESO 243$-$49 HLX$-$1.
There are two main types of search strategy used:
(1) exploring the luminous tail of the ULX luminosity distribution \citep[e.g.][]{sutton12}, 
and (2) studying the low mass end of SMBH mass distribution by using various scaling relations in dwarf galaxies \citep[e.g.][]{greene04}.
The first class of searches is usually based on ULX catalogs such as \citet{liu05} and \citet{walton11}.
Although they provide a meaningful list of sources for deeper studies, they are however limited by the narrow selection of ULX host galaxies, where for example ULXs are sought using a cross-match between the {\it XMM-Newton} source catalog and nearby galaxy catalogs, i.e. the Third Reference Catalogue of Bright Galaxies (RC3) \citep{vaucouleurs91} containing $\simeq 23 000$ galaxies.
The RC3 is well covered by X-ray observations and hence is a good input catalog for ULX searches. But for our purposes it contains a limited total solid angle and limited search volume (especially important for such rare objects as accreting IMBHs) compared to modern large galaxy surveys.

To overcome these limitations we propose an extended approach to select HLX candidates:
we cross-correlate the {\it XMM-Newton} source catalog \citep{rosen15} with one of the largest available galaxy redshift surveys,
the Sloan Digital Sky Survey Data Release 12 \citep{alam15} containing data on more than 3 million galaxies with known distances, using a special match condition.
We then apply several filter criteria to discard most of the known contaminating object classes based on their broadband spectral energy distribution (SED) properties and X-ray spectral features.
In this paper we present this search method and a selection of 2 candidate HLXs representing a snapshot of its early results. 

The paper is organised as follows. 
In \sref{sec:selection} we describe our selection methodology and how we address the most important issues of foreground and background contamination of the candidate sample with conventional object classes such as AGN, stars, etc.
In \sref{sec:analysis} we present the details of data analysis.
In \sref{sec:cand} we present 2 candidates and their observed properties.
\sref{sec:discussion} is devoted to the discussion on the candidates.
We give our conclusions in \sref{sec:conclusion}.

\section{Sample selection}
\label{sec:selection}

The goal of this study is to find extragalactic off-nuclear X-ray sources in the HLX X-ray luminosity range $10^{41} < L_{\rm X} < 10^{44}$\,\ergs, similar to ESO 243$-$49 HLX$-$1.
This source is observed as a point X-ray object 8\arcsec\, away from the nucleus of the host spiral galaxy and it exhibits spectral states similar to accreting StMBH:
a soft power-law with index of around 3 observed once and tentatively classified as the very high state, a harder power-law with a slope $\Gamma$=2.2 and a blackbody disk component with temperature $T \simeq 0.2$~keV in the high soft state, and a simple hard power-law with a power-law index in the range 1.4--2.1 in the low hard state \citep{godet09,servillat11,godet12}.
Its observed luminosity changes by almost two orders of magnitude between the low and the high states, from $2 \times 10^{40}\,\ergs$ to $\simeq 1 \times 10^{42}\,\ergs$ in the 0.2--10~keV band.

The observed properties of ESO 243$-$49 HLX$-$1 define our main search criteria:
we look for X-ray sources in the outskirts of galaxies whose distance is known, in order to be able to convert the observed flux to an estimate of the X-ray luminosity.
At the same time we apply a luminosity filter and select X-ray sources with inferred luminosities in the range $10^{41} < L_{\rm X} < 10^{44}$\,\ergs,
which translates to the mass of few hundred to few $\times 10^5$\,\msun for a black hole accreting at the Eddington limit.
We do not impose, however, any constraints on X-ray spectrum or variability of selected X-ray sources because it is uncertain if these properties are unique for ESO 243$-$49 HLX$-$1 or not.

In this study we chose to work with the {\it XMM-Newton} source catalog as an input list of X-ray sources mainly because whilst {\it XMM-Newton} has only fair spatial resolution with half energy width of 15\arcsec, it has excellent sensitivity over a large field of view and a broad energy range.
We used its recent release 3XMM-DR5 \citep{rosen15} which is the largest existing catalog of X-ray sources available publicly.

Throughout the paper we refer to source properties as they are given in the official distributed 3XMM-DR5 catalog files.
As the first step of our selection procedure we applied a filter to the 3XMM-DR5 sources.
We selected only point sources (those with {\tt SC\_EXT\_ML} = 0 in the catalog) with a detection significance of more than 8 ({\tt SC\_DET\_ML} $> 8$), which corresponds to a detection significance of $\approx 3.4\sigma$ in order to discard spurious objects.
Point source tag in the catalog is set when an attempt to fit PSF convolved with $\beta$-model for extended objects results in an extent parameter of less than 6\arcsec\footnote{More details on this procedure is available in the {\it XMM-Newton} catalog pipeline documentation at \url{http://xmmssc-www.star.le.ac.uk/Catalogue/2XMM/SSC-AIP-TN-003.ps}}.
This filter yielded 309,327 sources out of an initial list of 396,910 unique sources contained in the catalog.

\begin{figure}
\plotone{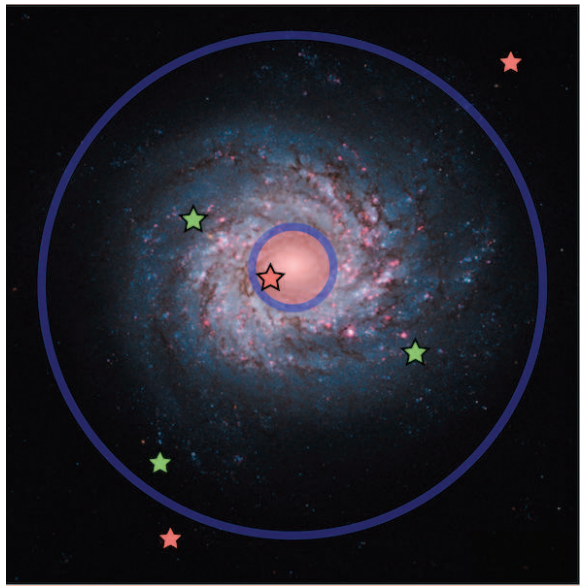}
\caption{Illustration of the off-nuclear cross-match method. Optical (represented by a galaxy) and point X-ray sources (represented by small green and red stars) are considered matched if X-ray source lie within 2 Petrosian radii from the galaxy center (outer circle) but further than the inner circle which accounts for galaxy nucleus. Positively matched X-ray objects are denoted with green stars, negative matches -- with red ones. Exclusion of the nucleus is required to discard contaminating AGN. \label{fig_xmatch}}
\end{figure}

Then we performed an off-nuclear cross-match (see Fig.~\ref{fig_xmatch} for the illustration of its concepts) of the X-ray source list with the spectroscopic sample of the SDSS DR12 catalog \citep{alam15}.

An X-ray source was considered for further analysis if its angular distance from the center of a galaxy was between the inner radius $r_{\rm in}$ and outer radius $r_{\rm out}$ calculated individually for each potential galaxy/X-ray source pair. 
The {\em inner radius} $r_{\rm in}$ was used to exclude X-ray sources that were too close to the galaxy nucleus, as it would not be possible to distinguish them from an AGN or to exclude contamination from it. 
If {\tt SC\_POSERR} is the $1\sigma$ position error of the X-ray source (its median value in our catalog sample is 1.4\arcsec), we chose this radius to be {\tt 3 * SC\_POSERR} but not less than 5\arcsec.
This minimum $r_{\rm in}$ limit was chosen after \citet{walton11} reasoning with the correction that the overall 3XMM-DR5 astrometry precision is significantly better than the one of 2XMMi-DR3 used by them \citep{rosen15}.

The {\em outer radius} $r_{\rm out}$ was chosen in order to have a reasonable chance that the source was associated with the galaxy and not be a background/foreground source.
We used an outer radius of {\tt 2 * petroRad\_r + 3 * SC\_POSERR} where {\tt petroRad\_r} is the modified Petrosian radius \citep{petrosian76} of a galaxy in the $r$ band adopted in the SDSS.
The Petrosian radius is defined using the shape of the azimuthally averaged light profile and SDSS uses 2 Petrosian radii to recover all the flux of a galaxy ({\tt petroMag}).
We use the Petrosian radius in this study instead of the isophotal diameter $D_{25}$
(e.g. given for galaxies from the RC3 catalog) which is more common for ULX searches, because it provides more uniform galaxy flux and extent measures across possible galaxy brightness profiles, independently from brightness profile fits.

It is not easy in general to characterise the ratio between the Petrosian radius and the $D_{25}$ as it depends on the brightness profile of a galaxy,
but in some samples Petrosian diameters are generally found to be smaller than the projected major axis of a galaxy at the 25 mag~arcsec$^{-2}$ isophotal level.

As a result of the off-nuclear cross-match we obtained a list of X-ray source/galaxy pairs.
Therefore, we were able to compute the X-ray luminosity of each source under the assumption that it was associated with the matched galaxy with known distance.
Having done this, we then selected only those X-ray sources that satisfied our luminosity criterion to be considered as HLX candidates: $10^{41} < L_{\rm X} < 10^{44}\,\ergs$. 
At this step we also restricted our study to the objects with distances less than 800~Mpc which corresponds to the limiting sensitivity of the {\it XMM-Newton} for objects with luminosity at the fainter end of the studied range.
This intermediate list had 373 objects.

Obviously, by construction the selected intermediate sample was prone to contain superpositions of either background or foreground X-ray objects onto optical galaxies.
Since we are interested only in candidate hyperluminous X-ray sources, throughout this paper we refer to the conventional types of sources (which are not associated with the matched galaxies as we assume by default) such as active galactic nuclei (AGN), X-ray active stars, cataclysmic variables (CVs), etc, as contaminants.
Main contaminant object types can be {\it background} AGN, {\it background} BL Lac objects, {\it background} starburst galaxies, Galactic {\it foreground} compact objects, Galactic {\it foreground} stars.

We tried to address the problem of cleaning our HLX candidates from the contaminants by using broadband spectral energy distribution (SED) properties of different object types, in particular the X-ray-to-optical flux ratio $f_{\rm X}/f_{\rm opt}$.
To compute it we use the X-ray flux in the standard {\it XMM-Newton} catalog band 0.2--12~keV and the optical flux in the SDSS $r$ band throughout this paper.
We note that $f_{\rm X}/f_{\rm opt}$ ratios from the literature are highly inhomogeneous with respect to the actual bands used and further in the paper we either tried to bring them to the same reference system or indicate that the estimate is uncertain.
The X-ray-to-optical flux ratio $f_{\rm X}/f_{\rm opt}$ of ESO 243$-$49 HLX$-$1 is usually large, around 100, judging from optical and X-ray data from \citet{webb14}, though the amplitude of its variability is a factor of $\approx 50$ in the X-ray and a factor of several in the optical.
For ULXs typical $f_{\rm X}/f_{\rm opt}$ values are $10^2 - 10^3$ \citep{tao11}.
The most common contaminating object types, AGN and foreground stars, have instead low X-ray-to-optical flux ratios: it is rarely more than 10 for AGN and always less than 0.1 for stars \citep{lin12}.
However, distant obscured AGN can reach extreme X-ray-to-optical flux ratios of $10^2$ and much more \citep{mignoli04,bauer04}.
This population of objects with extreme $f_{\rm X}/f_{\rm opt}$ values starts to emerge at fluxes roughly below few $\times 10^{-14}$\,\ergscm (e.g. \citet{comastri02}, see also fig.~7 of \citet{bauer04}, fig.~1 of \citet{mignoli04}).
To minimize their contamination we imposed X-ray flux limit of $f_{\rm X} > 5 \times 10^{-14}$\,\ergscm for our intermediate selection which gave us the list of 98 HLX candidates we later refer to as the base sample.

The full details of our filtering and cross-match method described to this point, including the exact queries we used to construct our samples, are given in the Appendix~\ref{sec:appendix_details}.
The intermediate and base tables of our HLX candidates can be easily obtained from the SDSS CasJobs service with these queries.

\subsection{Estimate of contaminating sources contribution}
\label{sec:contaminants}

It is possible to estimate the contribution of unknown contaminants to our base sample.
We follow the procedure described by \citet{walton11} with minor modification: we do not account for non-uniform sensitivity of the {\it XMM-Newton} observations which would only decrease more realistic contamination estimate. Hence we present here its upper limit.
In total there are 33,879 extragalactic objects from the SDSS DR12 spectroscopic sample which were observed in the 3XMM-DR5 catalog set of observations and were fed as an input to our off-nuclear cross-match method. 
To summarize, these are galaxies with distance less than 800~Mpc and Petrosian radius larger than 2.5~arcsec -- the later would have zero area after applying our off-nuclear criterion. 
The footprint of 3XMM-DR5 observations is defined by its preliminary multi-order coverage (MOC) map computed from the combined images\footnote{Available from \url{http://xmm-catalog.irap.omp.eu/links}.}.
This is the simplest footprint estimate available so far and it does not take into account exposure maps and variable observation sensitivity, so the number of SDSS galaxies inside it may be considered as an upper limit on number of galaxies which we inspected with our off-nuclear cross-match technique.
In the same way as \citet{walton11} for each individual galaxy falling in the 3XMM-DR5 footprint we then compute the number of contaminating sources in its solid angle (with the nuclear region subtracted) using the log N -- log S relation for resolved X-ray background sources in the hard (2--10~keV) X-ray band from \citet{moretti03}. 
We use the following formula to compute individual sensitivity limit for each galaxy: 

\begin{equation}
S = \alpha \cdot {\rm max}(f_{\rm X_{lim}}, \frac{L_{\rm X_{min}}}{4 \pi d^2 })
\end{equation}

in \ergscm, where $\alpha = 0.7$ is a factor from \citet{walton11} to account for the average fraction of the flux observed in the hard band (2--12~keV) compared to the full {\it XMM-Newton} 0.2--12~keV band for an assumed background source spectrum, $f_{\rm X_{lim}} = 5 \times 10^{-14}$\,\ergscm is the minimum X-ray flux limit we adopted when constructing our sample, $L_{\rm X_{min}} = 10^{41}$\,\ergscm is the lower limit of our search luminosity range, $d$ is luminosity distance computed from galaxy redshift using standard cosmology adopted by the SDSS CasJobs service.
This individual limiting sensitivity $S$ value is then subsituted to log N -- log S expression from \citet{moretti03} to get a number of expected background sources above this sensitivity $N(>S)$ in the searched area of a given galaxy.
We do not expect any major corrections to our contamination estimate due to the slight differences in energy range between the hard band of \citet{moretti03} and that of the {\it XMM-Newton}.

Finally, after summing individual values of number of background sources computed for each galaxy sensitivity limit and solid angle we get total value of $71 \pm 11$ contaminants we expect to have in our sample of 98 sources.
The quoted uncertainty corresponds to the 15~per~cent uncertainty of the hard band number counts from \citet{moretti03}.

\citet{moretti03} used several wide-field and pencil-beam X-ray surveys to compute their log N -- log S relation. 
They performed a detailed comparison of their results with the literature and found that their cosmic X-ray background (CXB) model agrees with other studies within the uncertainties, i.e. there is no evidence that their study is biased due to the cosmic variance or other reasons.
Nevertheless we verify our contamination estimate using other published log N -- log S functions.
We note that \citet{moretti03} function provides systematically larger source number counts than that of {\it Chandra}-COSMOS and {\it XMM}-COSMOS surveys \citep[][see fig.~9]{elvis09} and therefore our contaminants estimate would only be lower based on these log N -- log S data.
\citet{georgakakis08} combined several surveys such as CDF-N, CDF-S, EGS and XBOOTES to derive soft (0.5--2.0~keV), hard (2--10~keV) and total (0.5--10~keV) band log N -- log S relations.
Using hard band log N -- log S from \citet{georgakakis08} (i.e. setting $\alpha =0.7$ in Equation 1 as above) yields 59 contaminants in our sample.
With their total band relation (also setting $\alpha =1$ in Equation 1) one gets 73 contaminants.
Both methods agree with \citet{moretti03} in a statistical sense and we shall use its results throughout this paper.

With this simple approach we do not take into account CCD chips gaps and searched galaxies area overlap both reducing the total solid angle of our survey and the flux extinction that will be suffered by background sources due to the gas and dust in the galaxies these sources are falsely associated with.
Probably more important, we do not correct for the varying sensitivity of observations like \citet{walton11} did because for the very recent release of 3XMM-DR5 catalog there is no upper limit server available yet to the community like Flix\footnote{\url{www.ledas.ac.uk/flix/flix3.html}} which is available for older catalog releases.
It is clear that all the effects above and the shallow X-ray observations of a fraction of 33,879 input galaxies with detection limits above our $S$ estimates for them would only decrease our contamination estimate, so fractional contamination of $72 \pm 11$~per~cent should be considered as a conservative upper limit.

\subsection{Final sample}
\label{sec:final_sample}

Upper limit for fractional contamination of 72~per~cent allows us to expect that in our base sample of 98 objects at least few are good HLX candidates.
Ideally, one needs to obtain spectroscopy to confirm that the distances to the candidate HLXs are indeed those of the supposed host galaxies. 
However, getting such a confirmation for all 98 objects is very expensive observationally, so in this paper we present a smaller subset of the HLX candidates for which we find evidence in the public data available through the Virtual Observatory (VO) that their observed properties differ from those of the main contaminating object type, AGN.

We use several criteria based on the SED analysis of the objects from our base sample.
As we mentioned earlier, we limited our selection to the sources with $f_{\rm} > 5 \times 10^{-14}$\,\ergscm to filter out population of AGN with extreme X-ray-to-optical flux ratios.
However, not only distant intrinsically obscured AGN below this flux level have high X-ray-to-optical ratios.
Conventional AGN with intrinsic $f_{\rm X}/f_{\rm opt} = 1$ seen through significant absorption material (e.g. through a foreground galaxy) can have very large observed X-ray-to-optical flux ratio.
An example of such case is a bright ULX candidate with observed $f_{\rm X}/f_{\rm opt} \simeq 100$ discovered by \citet{miniutti06} which was later proved by \citet{dadina13} to be background absorbed AGN with intrinsic $f_{\rm X}/f_{\rm opt} \simeq 6$.
More recently, \citet{sutton15} has obtained optical spectrum of another HLX candidate 2XMM J134404.1$–$271410 with $f_{\rm X}/f_{\rm opt} \simeq 50$ which turned out to be quasar at redshift $\simeq 2.84$ seen through nearby galaxy IC~4320.
At the same time there are a several ULXs known to have values of $f_{\rm X}/f_{\rm opt} \lesssim 100$ because e.g. they are embedded in optically bright HII regions in their host galaxy or show significant variability \citep{tao11,heida13}.

One possible way out from this significant overlap of $f_{\rm X}/f_{\rm opt}$ of AGN and compact accreting objetcs is to compare X-ray and near-infrared (NIR) fluxes as NIR is less sensitive to the line of sight absorption than optical light.
Below we refer to the flux in the $K$ band as to the NIR flux.
It is easy to estimate that a line of sight absorption of $n_{\rm H} = 10^{22}$~cm$^{-2}$ with normal gas-to-dust ratio for a typical AGN increases its intrinsic X-ray-to-optical flux ratio by a factor of $\simeq 70$, whereas for $f_{\rm X}/f_{\rm NIR}$ it is only a factor of 1.05.
Moreover, extreme X-ray-to-optical flux ratio objects such as EROs (extremely red objects) have very decent $f_{\rm X}/f_{\rm NIR} \lesssim 25$ while showing X-ray-to-optical flux ratios of hundreds and even thousands \citep{mignoli04}.
In the sample of ULX candidates observed in the NIR by \citet{heida14} AGN seen through the foreground galaxies possess moderate $f_{\rm X}/f_{\rm NIR} \lesssim 10$ while ULXs have such a ratio greater than $\simeq 40$ for e.g. XMMU~J024323.5+372038 in NGC~1058 and RX~J004722.4-252051 in NGC~253 to $\simeq 2000$ for Holmberg II X$-$1.

Based on these considerations we decided to start studying our base sample from sources that could satisfy $f_{\rm X}/f_{\rm opt} > 100$ or $f_{\rm X}/f_{\rm NIR} > 40$ criteria based on data existing in the Virtual Observatory without obtaining new observations.
We note however, that sources from our base sample which do not satisfy these criteria are still worth further investigation.

As the first step of this selection of optically/NIR quiet objects, we apply the following criterion based on the SDSS data.
The non-detection of an optical source within the 3$\sigma$ X-ray error circle up to the limit of the SDSS (22.2~AB~mag in band $r$), gives one lower limit on this ratio $f_{\rm X}/f_{\rm opt} \gtrsim 10$ for the minimum X-ray flux of our sample ($f_{\rm X} = 5 \times 10^{-14}$\,\ergscm in the standard {\it XMM-Newton} energy band 0.2--12~keV).
We hence filtered from the base sample those sources that have an optical object in the SDSS DR12 photometric database within their $3\sigma$ X-ray positional uncertainty (61 objects). 
These sources with optical counterparts thus having $f_{\rm X}/f_{\rm opt} \lesssim 10$ are likely to be reasonably distant AGN that dominate the resolved X-ray background at fluxes $f_{\rm X} > 5 \times 10^{-14}$\,\ergscm. 
We note that their number is in good agreement with our contamination estimate.
This test passage left 37 optically faint sources for further inspection.
Known HLX candidates such as ESO 243$-$49~HLX$-$1 and M82~X$-$1 are not included in this sample due to incompleteness of the optical redshift survey we used i.e. SDSS DR12 does not cover the full sky.
When testing this method using other redshift databases such as HyperLEDA in the input, we were able to recover these HLX candidates.
New HLX candidates from other redshift databases will be presented in the future papers.

To strengthen obtained $f_{\rm X}/f_{\rm opt} \gtrsim 10$ limit we queried the archive of the Canadian Astronomical Data Centre (CADC) which hosts deep imaging data from the Hubble Space Telescope, CFHT Megacam and WIRCam and other instruments.
We used Virtual Observatory's Table Access Protocol \citep{dowler11} service at the CADC to find and retreive cut out images of 5 out of 37 objects from the subset above.
We then analysed their deep optical images. 
For one object (XMM1226+12, see below) we find evidence that it satisfies chosen constraint $f_{\rm X}/f_{\rm opt} > 100$.
Other 4 objects either have relatively bright optical counterparts and $f_{\rm X}/f_{\rm opt} \simeq 20...50$, or do not have optical counterparts but project on diffuse host galaxy light which only allows us to obtain insufficiently strict lower limit of $f_{\rm X}/f_{\rm opt} \gtrsim 30$.
Nevertheless, we found it reasonable to start observational campaign to follow up these candidates at optical/NIR wavelengths in order to obtain decisive evidence on association of X-ray sources counterparts with assumed host galaxies and/or improve $f_{\rm X}/f_{\rm opt}$ and $f_{\rm X}/f_{\rm NIR}$ constraints available from the Virtual Observatory.

\begin{figure*}
\epsscale{2.0}
\plotone{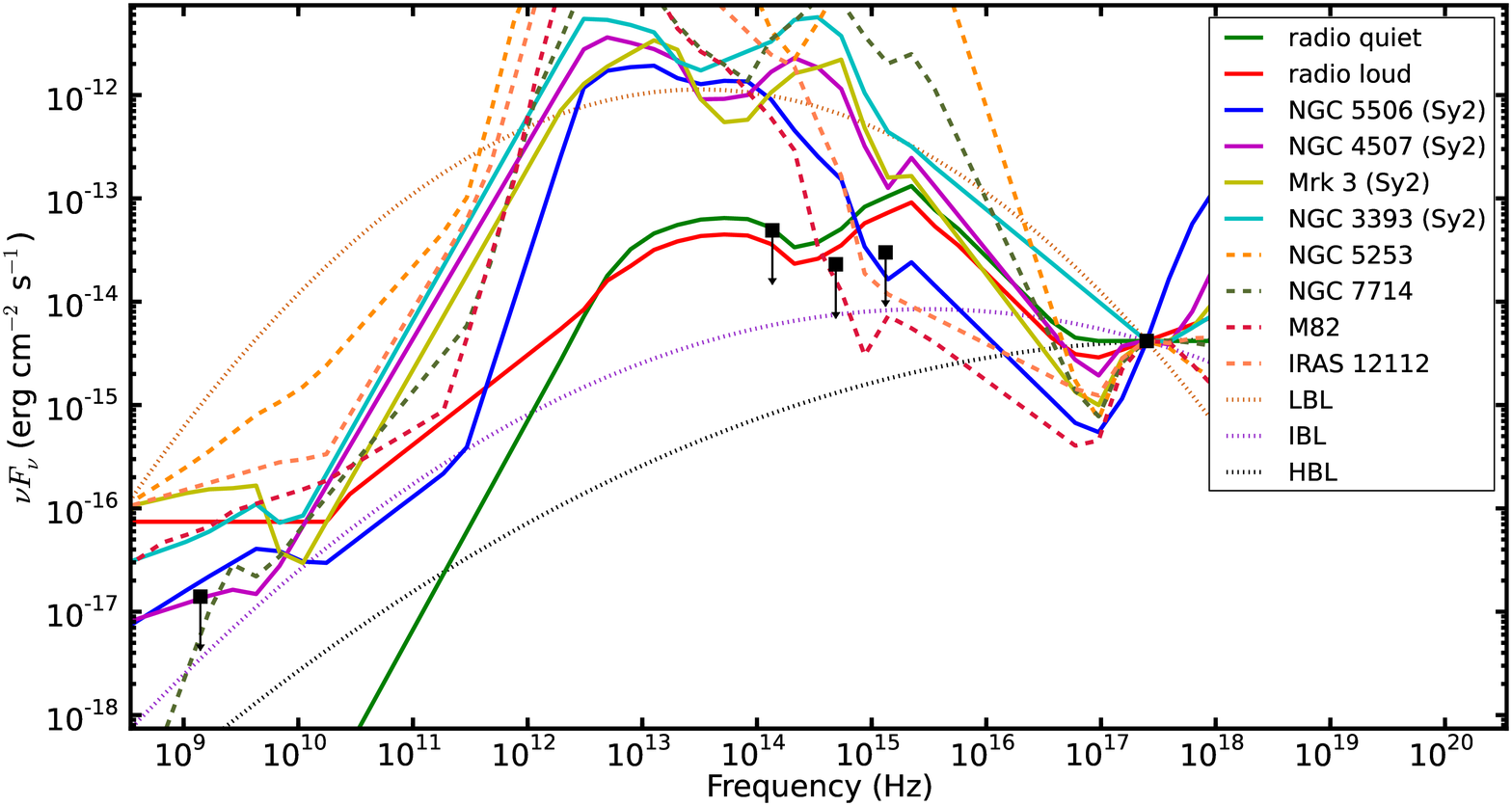}
\caption{X-ray-to-radio spectral energy distributions of AGN (solid lines), starburst galaxies (dashed lines) and BL Lac objects (dotted lines) normalized to match the minimum X-ray flux of our base sample. 
Approximate flux limits for VLA FIRST in the radio (1 mJy at 1.4 GHz), UKIDSS Large Area Survey in the near-infrared ($K > 18.1$ Vega mag), SDSS photometry in the optical ($r > 22.2$ AB mag), {\it GALEX} Medium Imaging Survey ($m_{AB} > 23$) are shown as squares with arrows from left to right, respectively. Minimum X-ray flux of our base sample ($5 \times 10^{-14}$\,\ergscm from 0.2 to 12~keV which approximately translates to $\nu F_\nu = 4 \times 10^{-15}$\,\ergscm at 1~keV) is shown as a square on the right. It is evident that with the data from public surveys alone one can claim that neither AGN nor starburst galaxies nor low-energy-peaked BL Lacs (LBL) satisfy the observed X-ray detection and radio/NIR/optical/UV non-detection constraints simultaneously. 
Intermediate- and high-energy-peaked BL Lacs can explain observed SED properties, but these can usually be eliminated at the in-depth analysis stage based on their hard X-ray spectra and/or deeper optical/NIR constraints. 
SED data are from \citet{ruiz10} except for the BL Lacs derived from \citet{nieppola06}. 
\label{fig_seds}}
\end{figure*}

We further developed this idea of discarding other contaminating types of objects by using their broadband SED properties.
In Figure~\ref{fig_seds} we plot observed X-ray-to-radio SEDs of several contaminating object types: AGN, starburst galaxies \citep{ruiz10} and BL Lacs \citep{nieppola06} normalized to the minimum X-ray flux of sources from our sample $f_{\rm X} = 5 \times 10^{-14}$\,\ergscm.
When adding upper limits of several other publicly available surveys from radio to near-infrared (NIR) to ultra-violet (UV) domains,
it becomes immediately clear that the detection of an object in the X-ray together with a non-detection at other wavelengths in several existing public surveys permits us to filter out all contaminating object classes listed above except for the rare high-energy-peaked blazars (HBL),
which however can be discarded at later stages when studying the X-ray spectra, as these possess characteristic hard and flat spectra.

To make this test we involved surveys at other wavelengths, namely the UKIRT Infrared Deep Sky Survey (UKIDSS) in the near-infrared \citep{lawrence07}, the Galaxy Evolution Explorer ({\it GALEX}) in the ultra-violet \citep{martin05} and the VLA FIRST \citep{becker95} in the radio, all easily accessible from many sites through the VO infrastructure.
We performed a cross-match within $3\sigma$ positional uncertainties plus a typical survey resolution from the X-ray coordinates with the UKIDSS Data Release 10+ (1\arcsec\, resolution), the {\it GALEX} Data Release 6 (3\arcsec\, resolution) and the VLA FIRST (5\arcsec\, resolution) catalogs and also inspected their imaging data.
This was done with the help of the {\sc topcat} VO-enabled table processing software \citep{taylor05} and {\sc cds aladin} VO image browser \citep{bonnarel00} and its results were used in the next step.
We note that at the typical magnitude limit of UKIDSS surveys $K = 18.3$ Vega mag, non-detection of a NIR counterpart translates to $f_{\rm X}/f_{\rm NIR} \gtrsim 15$ for the minimum X-ray flux of our base sample $f_{\rm X} = 5 \times 10^{-14}$\,\ergscm.
After this test we found one object (XMM0838+24, see below) that has required $f_{\rm X}/f_{\rm NIR} > 40$.

In this paper we present these 2 HLX candidates that satisfied chosen constraints and passed the SED test above.

\begin{figure*}
\epsscale{2.2}
\plotone{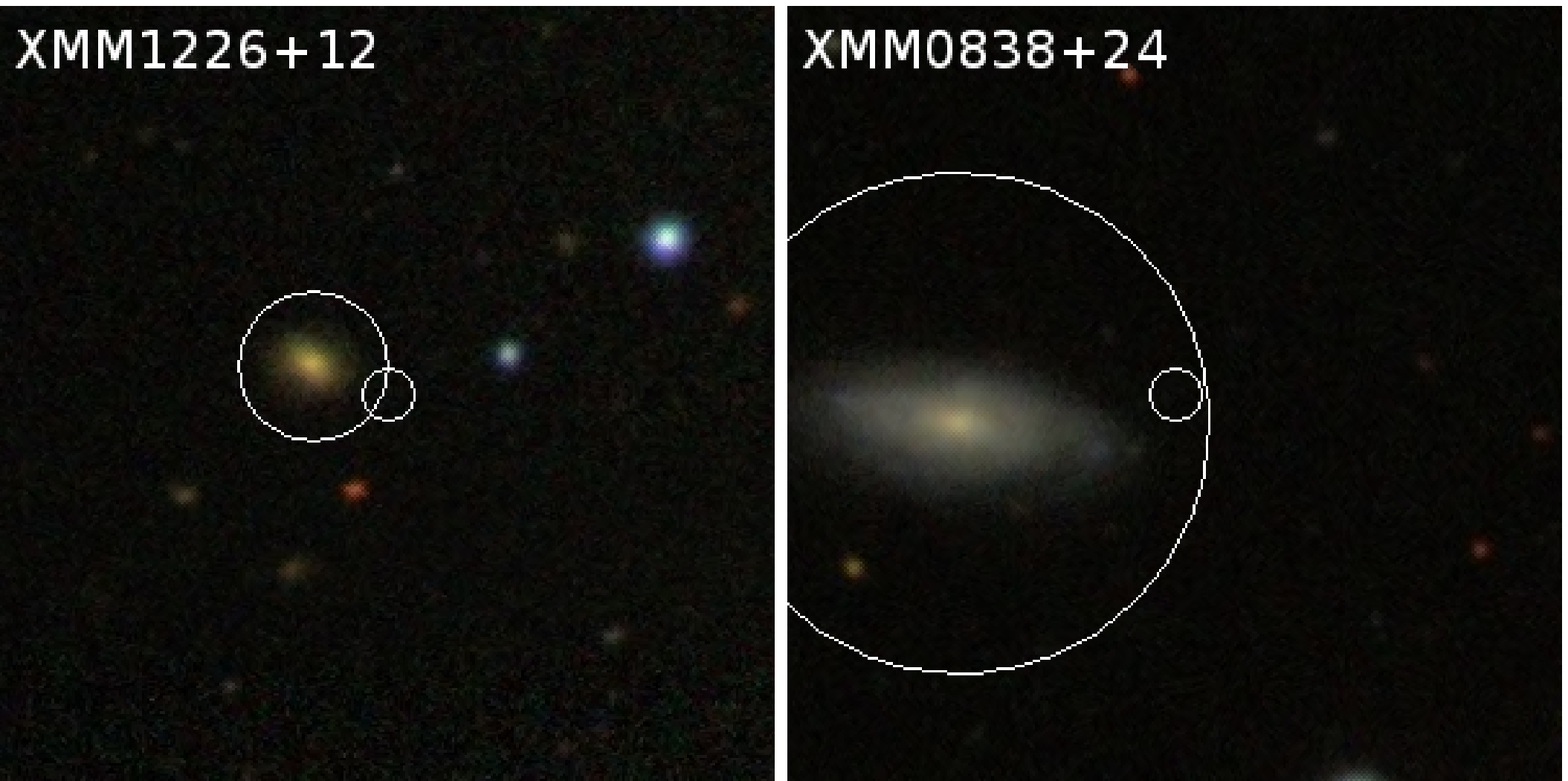}
\caption{Finding charts in the optical of the selected HLX candidates and their assumed host galaxies from the SDSS DR12 survey. 
Each chart has 1\arcmin\, side; north is up and east is to the left. Smaller circles in the centers of each image correspond to 3$\sigma$ positional uncertainty of an object from the 3XMM-DR5 catalog. 
Larger circles have radii of $2 \times$ Petrosian radius of an assumed host galaxy and represent one of the measures of optical confines of a galaxy. 
Each thumbnail image is labelled with the corresponding object identifier in the top left corner. \label{fig_finding_charts}}
\end{figure*}

\section{Data analysis}
\label{sec:analysis}

In this paper we chose to work with science ready data conveniently available through the Virtual Observatory framework.
X-ray data in present study were taken from {\it XMM-Newton}'s 3XMM-DR5 catalog \citep{rosen15} accessible from the catalog website\footnote{\url{http://xmm-catalog.irap.omp.eu}} and {\it Chandra} Source Catalog Release 1.1 \citep{evans10} which we queried through the CDS VizieR service.
Other tabular catalog data were accessed in a similar way using VO tools such as {\sc cds aladin} and {\sc topcat}.

We performed photometric and astrometric analysis of the flux calibrated optical and NIR images obtained through the CADC.
For this purpose we used SExtractor software \citep{bertin96} coupled with PSFEx \citep{bertin11}, versions 2.19.5 and 3.17.1 respectively.
We used SExtractor's built-in star/galaxy classifiers {\tt CLASS\_STAR} and {\tt SPREAD\_MODEL} and depending on their results took either PSF or Petrosian magnitude as an optical/NIR flux estimate.
For photometric upper limit estimates we used either measured magnitudes of stars at signal-to-noise ratio of 5, respective instrument exposure time calculator estimate, or reduction pipeline values from FITS headers, whichever is brighter.

We used the following approach to compute $f_{\rm X}/f_{\rm opt}$ and $f_{\rm X}/f_{\rm NIR}$ ratios.
X-ray flux $f_{\rm X}$ was taken from the 3XMM-DR5 unless indicated other.
For optical flux $f_{\rm opt}$ we first converted literature or this paper measurements to AB magnitudes and obtained spectral flux density estimate.
Same procedure was followed to obtain NIR spectral flux density.
Then assuming it does not change much within the bandwidth of SDSS $r$ or UKIDSS $K$ filter we computed total flux within a bandpass and used it to compute the ratio with given $f_{\rm X}$.

\section{Results}
\label{sec:cand}

Positional information such as X-ray coordinates and their uncertainties on selected HLX candidates is presented in Table~\ref{tab_candidates_posinfo}.
We give the main properties of the candidates in Table~\ref{tab_candidates}.
There we also present some parameters of the sources derived under the assumption that they are associated with the optical galaxies.
Available properties of the assumed host galaxies are listed in Table~\ref{tab_galaxies}. 
In this Section we present the facts we have on the individual candidates.

\begin{deluxetable}{ccccccc}
\tablecaption{Positional information on HLX candidates.\label{tab_candidates_posinfo}}
\tablewidth{0pt}
\tablehead{
\colhead{Source} & \colhead{Short name} & \colhead{RA} & \colhead{Dec} & \colhead{Pos. err.} & \colhead{$l$} & \colhead{$b$} \\
\colhead{ID} & & \colhead{J2000} & \colhead{J2000} & \colhead{(arcsec)} & \colhead{(deg)} & \colhead{(deg)}
}
\startdata

201082602010056 & XMM1226+12 & 12:26:47.76 & +12:55:04.2 & 1.0 & 279.6357 & 74.6640 \\
203022602010004 & XMM0838+24 & 08:38:39.47 & +24:53:09.5 & 0.9 & 199.8610 & 33.7771 \\

\enddata
\tablecomments{Source IDs are given from the {\tt srcid} column in the 3XMM-DR5 catalog. Positional error is the $1\sigma$ coordinate uncertainty. $l$ and $b$ are the Galactic longitude and latitude respectively.}
\end{deluxetable}

\begin{deluxetable}{ccccccccc}
\tablecaption{HLX candidates and their some characteristics.\label{tab_candidates}}
\tablewidth{0pt}
\tablehead{
\colhead{Source} & \colhead{Flux} & \colhead{$f_{\rm X}/f_{\rm opt}$} & \colhead{Luminosity} & \colhead{Mass} & \colhead{Distance} & \colhead{Separation} & \colhead{Pos. err.} \\
\colhead{ID} & \colhead{(\ergscm)} & \colhead{} & \colhead{(\ergs)} & \colhead{($\msun$)} & \colhead{(Mpc)} & \colhead{(arcsec)} & \colhead{(arcsec)}
}
\startdata
201082602010056  & $5.7\times10^{-14}$ &   $\simeq 200$ (opt)    & $4.0\times10^{42}$ & $\simeq 31000$   & 754  & 9.1 & 1.0 \\
203022602010004  & $1.4\times10^{-13}$ &   $\gtrsim70$ (NIR)     & $2.7\times10^{41}$ & $\simeq 2100$   & 128  & 24.4 & 0.9 \\
\enddata
\tablecomments{Distances derived from the SDSS redshifts are given to a candidate host galaxy (see also Table~\ref{tab_galaxies} for other host galaxies properties).
The flux column gives the observed 0.2--12~keV flux which is taken from the 3XMM-DR5 catalog.
$f_{\rm X}/f_{\rm opt}$ is the ratio of the observed X-ray flux to the detected counterpart or detection limit flux in the optics or in the NIR. 
The luminosity is computed from the observed X-ray flux column with the assumption that the X-ray source is associated with the galaxy. 
The mass is calculated from the luminosity using simple Eddington luminosity scaling: $L_{\rm Edd} = 1.3 \times 10^{38} (M/\msun)\,\ergs$, rounded to two significant digits. 
The separation is given with respect to a host galaxy nucleus.
The positional error column is the total position uncertainty in arcseconds calculated by combining the statistical error and the systematic error, taken from the 3XMM-DR5 catalog ({\tt sc\_poserr} column).
}
\end{deluxetable}

\begin{deluxetable}{cccccc}
\tablecaption{Properties of assumed host galaxies for each of the selected HLX candidates.\label{tab_galaxies}}
\tablewidth{0pt}
\tablehead{
\colhead{Source} & \colhead{SDSS DR12 objid} & \colhead{Type} & \colhead{Petrosian radius} & \colhead{Redshift} \\
\colhead{ID} & \colhead{ID} & \colhead{} & \colhead{(arcsec)} & \colhead{}
}
\startdata

201082602010056   & 1237661813349351502 & $Sa (?)$  & 4.2   & 0.158 \\
203022602010004   & 1237664667887599821 & $Sc$      & 14.1 & 0.029 \\

\enddata
\tablecomments{Galaxies types are estimated from SDSS imaging and spectral information.
Question mark means that classification is not certain.}
\end{deluxetable}

\subsection{Source 201082602010056 (= XMM 1226\-+12)}

Source 3XMM J122647.7\-+125504\footnote{See full details on this source in the 3XMM-DR5 catalog at \url{http://xmm-catalog.irap.omp.eu/source/201082602010056}} ({\it XMM-Newt\-on} unique persistent catalog identifier 201082\-602010056, hereafter source XMM1226+12 for brevity; 
we follow the same nomenclature for another source) was observed on Jul 1, 2002 as part of observation 0108260201. 
The source is detected next to the {\it Sa} galaxy (see its thumbnail in the Fig.~\ref{fig_finding_charts}).
Judging from its SDSS optical spectrum (namely, its position on the Baldwin--Phillips--Terlevich (BPT) diagram \citep{baldwin81}) the assumed host galaxy itself is not active.
{\it XMM-Newton} catalog flux estimate is $f_{\rm X} = 5.9 \times 10^{-14}$\,\ergscm.

The source is present in \citet{liu11} catalog of X-ray sources from {\it Chandra} survey of nearby galaxies where it is associated with NGC~4406 (M86) galaxy at 16.8~Mpc which projects some 8.9\arcmin\, away.
M86 has $D_{25}$ major semi-axis of 4.5\arcmin\, according to \cite{vaucouleurs91}.
These {\it Chandra} data were obtained on Mar 9, 2005 and measured X-ray flux is $f_{\rm X} = 9.1 \times 10^{-15}$\,\ergscm in 0.3--8~keV band.
The source is also included in {\it Chandra} Source Catalog \citep{evans10} as CXO J122647.7+125505 and its ACIS aperture-corrected net flux from the same observation is estimated as $f_{\rm X} = 1.9 \times 10^{-14}$\,\ergscm in 0.5--7~keV band.

\begin{deluxetable}{ccccccc}
\tablecaption{CFHT Megacam observations of the XMM1226+12 field analysed in this study.\label{tab_megacam}}
\tablewidth{0pt}
\tablehead{
\colhead{Filter} & \colhead{Date} & \colhead{Exp. time} & \colhead{Mag. limit} & \colhead{Seeing} \\
 &  & \colhead{(s)} & \colhead{(AB mag)} & \colhead{(arcsec)}
}
\startdata

$r$ & Jan 17, 2005 & 1440 & 25.4 & 1.0  \\
$r$ & Apr 17, 2005 & 900 & 24.8 & 1.7  \\
$u$ & Apr 01, 2008 & 6984 & 27.0 & 0.8  \\
$r$ & Apr 03, 2008 & 4809 & 26.3 & 1.0  \\
\enddata
\end{deluxetable}

\begin{figure}
\epsscale{1.0}
\plotone{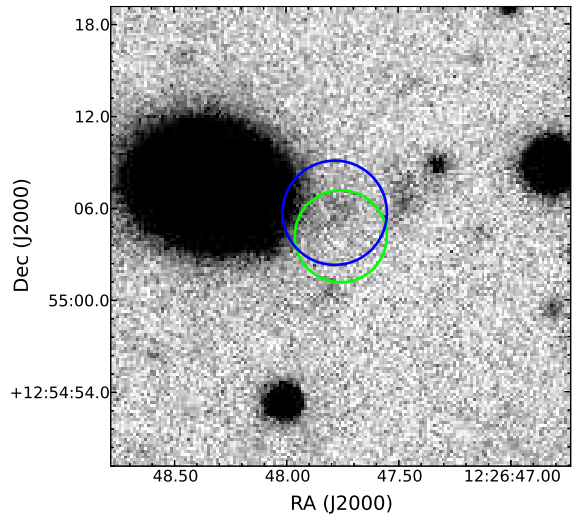}
\caption{Stacked CFHT Megacam image of the XMM1226+12 field in $r$ band obtained on Apr 3, 2008 with total exposure of 4809~sec with {\it XMM-Newton} (green) and {\it Chandra} (blue) $3\sigma$ positional uncertainties overplotted. North is up and east is to the left.
An extended optical source is visible inside combined positional uncertainty.
\label{fig_xmm1226}}
\end{figure}

CADC archive has wealth of imaging data in this field as it belongs to Virgo galaxy cluster.
We analysed several deep Megacam frames (see the observation log in Table~\ref{tab_megacam}).
On the deepest $r$ band exposure an optical source is visible within combined {\it XMM-Newton} and {\it Chandra} positional uncertainty (see Fig.~\ref{fig_xmm1226}) at RA=12:26:47.77, Dec=+12:55:06.0.
Further we refer to it as to the source A.
SExtractor classifies source A as a galaxy and it is indeed apparent that it is extended.
Petrosian AB magnitude estimate is $m_r = 24.44 \pm 0.13$ whereas if one assumes that source A is a blend, its brightest component has PSF modelled magnitude $m_r = 25.0 \pm 0.1$.
However, on the shallower $r$ band exposure obtained on Jan 17, 2005 source A is only marginally visible, it is classified as star and has Petrosian AB magnitude of $m_r = 26.0 \pm 0.4$ and PSF magnitude $m_r = 25.4 \pm 0.3$.
There is no source visible within combined {\it XMM-Newton} and {\it Chandra} positional uncertainty on Megacam image obtained on Apr 17, 2005 up to image limiting magnitude 24.8.
On the deep $u$ band frame with almost 7~ksec exposure time and better night conditions source A is marginally resolved into 3 components, two of which have PSF magnitude $m_u = 26.4 \pm 0.2$, and the third fainter one has $m_u = 27.0 \pm 0.3$.
Moreover, another source (source B) is detected at 2.5$\sigma$ level at coordinates RA=12:26:47.69, Dec=+12:55:04.0 in the south-west region of the combined X-ray positional uncertainty with $m_u = 27.2 \pm 0.4$.
It is also barely visible in the deepest $r$ band exposure so we consider source B to be a real object rather than a background fluctuation.

There is no counterpart visible in the UKIDSS Large Area Survey image of this field up to its limiting magnitude of 18.3 (Vega).

\subsection{Source 203022602010004 (= XMM 0838\-+24)}

Source 203022602010004\footnote{See full details on this source in the 3XMM-DR5 catalog at \url{http://xmm-catalog.irap.omp.eu/source/203022602010004}} (IAU designation 3XMM J083839.4\-+245309) was observed by the {\it XMM-Newton} twice, first time on Apr 9, 2005 as part of observation 0302260201 and then on Oct 9, 2005 during observation 0302260401.
For brevity in this paper we refer to it as XMM0838+24.
Both {\it XMM-Newton} observations of this source are consistent with each other within respective parameter uncertainties.
The source is situated some 24.4\arcsec\, away from the nucleus of the $Sc$ galaxy SDSS J083841.25\-+245306.3 (see its thumbnail in the Fig.~\ref{fig_finding_charts}).

XMM0838+24 is likely associated with the source 1SXPS J083839.3\-+245310 from the {\it Swift} source catalog \citep{evans14} which is situated 3.9~arcsec away from the {\it XMM-Newton} position.
Its 0.3--10~keV band flux from power-law fit is slightly larger than that of 3XMM-DR5 and estimated to be $2.9^{+1.6}_{-1.1} \times 10^{-13}$\,\ergscm.
{\it Swift} detected this source on 4 occasions between Oct 4, 2005 and Dec 17, 2007, and 3 times between Oct 10, 2005 and Nov 25, 2007 the source was not detected though $3\sigma$ upper limits of those observations were above its mean count rate determined from when it was detected.
The photon index for the source in 1SXPS catalog is rather hard, $\Gamma = 0.0^{+1.0}_{-0.7}$.

\begin{figure}
\epsscale{1.0}
\plotone{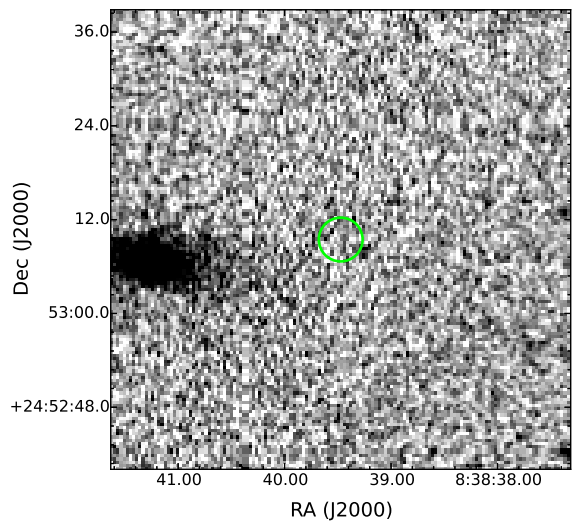}
\caption{$K$ band image of the XMM0838+24 field from UKIDSS Large Area Survey obtained on May 3, 2009 with 10~sec exposure time. 
North is up and east is to the left.
{\it XMM-Newton} $3\sigma$ positional uncertainty is overplotted.
A faint source is marginally visible just outside the {\it XMM-Newton} error circle.
\label{fig_xmm0838}}
\end{figure}

Despite that the cross-match of our base sample with UKIDSS DR10+ catalog did not give any result for this object, we inspected its field of view in $K$ band from the UKIDSS Large Area Survey (see Fig.~\ref{fig_xmm0838}).
There is a single object marginally visible at signal-to-noise ratio 3 just outside of the X-ray error circle at RA=08:38:39.28, Dec=+24:53:08.0 (3.1~arcsec separation whereas $3\sigma$ error circle radius is 3.0~arcsec).
Its Vega magnitude determined by PSF fitting is $K = 18.9 \pm 0.3$ which is deeper than typical UKIDSS detection limit because of the good quality of observing night.

\section{Discussion}
\label{sec:discussion}

\subsection{Possible nature of 2 HLX candidates} 

Although we cannot provide the exact nature of the selected candidates based on the existing data, we can describe their chances to be conventional non-HLX objects.
None of our sources could be a Galactic star: apart from our $f_{\rm X}/f_{\rm opt}$ criterion built into the selection procedure, these are easily recognized in optical SDSS images. 
Other foreground X-ray Galactic objects are less favored as well by the fact that the sample is constructed from the SDSS survey which intentionally spans high galactic latitudes where number densities of Galactic objects are low.
Both our sources lie above $+30^\circ$ (see Table~\ref{tab_candidates_posinfo}).
For example, only 34~per~cent of known CVs from the latest update of \citet{ritter03} catalog have a galactic latitude modulus larger than $33^\circ$ as this is the case for the source XMM0838+24.
From the complete flux-limited $f_{\rm X} > 10^{-14}$\,\ergscm survey of non-magnetic CVs (which comprise the majority of the local CV population) by \citet{pretorius12}, we can derive the total probability to have a single CV within the search solid angle of our sample to be as low as $\approx 10^{-4}$.
The typical flux of our sample $f_{\rm X} = 5 \times 10^{-14}$\,\ergscm translates to an observed luminosity of the order of $2.5 \times 10^{33}$\,\ergs at a maximum distance of 10~kpc that we can attribute to foreground Galactic objects at these latitudes.
This upper luminosity limit is far below the Eddington luminosity for any type of compact stellar mass object but still within a reasonable range for radiatively inefficient BH accretion.
Existing data therefore cannot completely rule out the Galactic nature of the sources, though by construction, high galactic latitudes make this hypothesis less probable, especially for XMM1226+12 situated at the latitude of almost $75^\circ$.
Below we discuss other non-extensive possibilities for each of two sources individually.

The expected host galaxy for the source XMM 1226\-+12 shows no sign of being an AGN, so it is unlikely that $L_{\rm X} \approx 4 \times 10^{42}\,\ergs$ (see Table~\ref{tab_candidates}) comes from the host galaxy itself, the origin of such X-ray emission in a normal galaxy is unclear.
{\it XMM-Newton} and {\it Chandra} observations of this source at different epoch with different flux estimates favor for its intrinsic X-ray variability even considering slight mismatch of X-ray bands.
This in turn makes its explanation as a superposition of fainter X-ray sources unlikely.
In the deepest $r$ band exposure of this field there is an extended source A within the combined positional uncertainty of X-ray coordinates, which is resolved in 3 components in $u$ band image, however, at low significance level.
If it was an inherently extended object it should be detectable in a 1440~sec $r$ image at the same flux level which is above the detection limit of that image.
If it is indeed a blend of several point objects like suggested by $u$ band image analysis, then we see the only one above the detection limit at 1440~sec exposure.
We hence assume that source A is a blend of several point-like objects and use its PSF magnitude in $r$ band to compute lower limit $f_{\rm X}/f_{\rm opt} \simeq 200$.
Unfortunately, the most shallow exposure of 900~sec was obtained during the night with worse than median weather conditions and has limiting magnitude that does not allow us to constrain optical variability of source A or its components.
Non-detection of the NIR counterpart in UKIDSS means its $f_{\rm X}/f_{\rm NIR} \gtrsim 17$.

Because of the high $f_{\rm X}/f_{\rm opt}$ value and $f_{\rm X}/f_{\rm NIR}$ lower limit this source is unlikely a background AGN.
One plausible explanation was suggested by \citet{liu11} who associated this X-ray object to M86 at 16.8~Mpc away.
On the sky plane XMM1226+12 is situated about twice the major semi-axis of $D_{25}$ ellipse of M86 away from its center at the positional angle of 30~degrees to the major axis, in the gap between M86 and NGC~4438 which is clearly seen as minimum on a radial distance distribution of M86 sources from \citet{liu11} catalog.
If XMM1226+12 nevertheless belongs to M86, it could be a low-mass X-ray binary (LMXB) with a projected distance of 43~kpc.
\citet{liu11} estimates its X-ray luminosity to be $3.1 \times 10^{38}$\,\ergs which is increased to $2.0 \times 10^{39}$\,\ergs if we consider observed {\it XMM-Newton} flux.
This would make this source one of the brightest black hole LMXBs in M86. 
However if this X-ray object is associated with either source A or source B, distance modulus of 31~mag to M86 implies its absolute optical magnitude $M_r \simeq -5...-6$, a bit brighter than one expects for LMXBs \citep{paradijs94,revnivtsev12}.
It is then valid to assume that this could be a LMXB situated in a low-mass globular cluster (GC), especially given that at this radial distance from M86 center there are $\simeq 3$ GCs per sq. arcmin \citep{rhode04} translating into decent probability of a chance superposition with our background galaxy of about 10~per~cent.
However colors of both brightest component of source A ($u - r \simeq 1.3$) and source B ($u - r \simeq 0.9$, considering deepest $r$ band magnitude limit) are too blue for a GC.
Indeed, in 2010 edition of \citet{harris96} catalog the bluest GCs have $u - r = 1.6...2.0$\footnote{We transformed their colors into AB system using \citet{jordi06} equations.}, in agreement with colors of dwarf spheroidal systems from \citet{chilingarian12} (see their fig.~C1).
Chance superposition probability of a rare blue faint GC in M86 with the background galaxy is neglible.
It is therefore unlikely that either the brightest component of source A or source B are field or GC LMXB though no claim can be made about fainter sources.
Data existing in the Virtual Observatory does not allow one to rule out the LMXB in M86 possibility with confidence.
More observations are required for definitive conclusions on the source's nature.

By sample construction due to the absence of radio, UV, optical and NIR counterparts the source XMM0838+24 is unlikely to be an active star, background AGN, starburst galaxy, and low/intermediate-/high-energy-peaked blazar (see Figure~\ref{fig_seds} but consider that detected X-ray flux is in any case above $1 \times 10^{13}$\,\ergscm). 

An estimate of $f_{\rm X}/f_{\rm NIR}$ depends if we assume marginally visible NIR source associated with the X-ray one.
If they are associated, then $f_{\rm X}/f_{\rm NIR} \simeq 70$.
Otherwise, this estimate becomes a lower limit.
In case we consider maximum X-ray flux from 1SXPS catalog and non-detection in NIR, the flux ratio of XMM0838+24 becomes $f_{\rm X}/f_{\rm NIR} \gtrsim 150$.

The derived luminosity for source XMM 0838\-+24 at 10~kpc is $\simeq 10^{33}$\,\ergscm. 
It is too low for a Galactic black hole binary undergoing near-Eddington accretion but may be plausible for an isolated black hole accreting from the interstellar medium (see e.g. \citet{maccarone05} and references therein). 
At the same time in this context indication that the observed X-ray spectrum of source XMM0838+24 does not coincide with predictions for various forms of radiatively inefficient accretion flows which are expected to have a power-law spectrum with a photon index 1.4--2.5 \citep[see e.g.][]{narayan98,ball01}.

High-resolution X-ray and deeper optical/NIR imaging and spectroscopic observations are required to shed light on the association of XMM0838\-+24 and its host galaxy and thus derive its X-ray luminosity.

\begin{figure}
\epsscale{.75}
\plotone{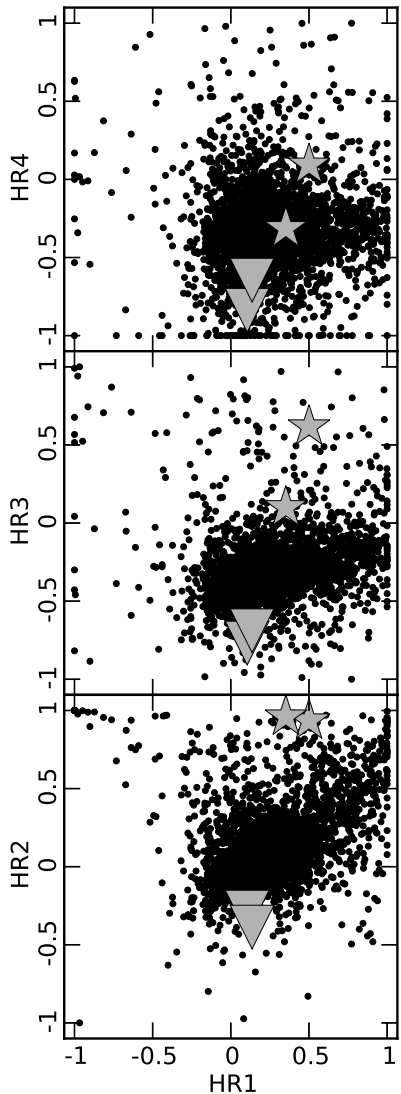}
\caption{The X-ray color-color diagrams for known AGN (from \citet{lin12}, black dots), selected HLX candidates (this paper, grey stars) and two spectral states of ESO 243$-$49 HLX$-$1, high luminosity consistent with steep power law and low luminosity hard power law states (grey triangles). 
X-ray colors are taken from the 3XMM-DR5 catalog and are defined as HR$_i = (f_{i + 1} - f_i) / (f_i + f_{i + 1})$, where $f_i$ is the X-ray flux in $i$-th {\it XMM-Newton} band (band 1 is 0.2--0.5~keV, band 2 is 0.5--1.0~keV, band 3 is 1.0--2.0~keV, band 4 is 2.0--4.5~keV, band 5 is 4.5--12~keV). 
As claimed by \citet{lin12} systematic uncertainties of these colors can reach 0.1 which is below symbol size for ESO 243$-$49 HLX$-$1 and HLX candidates. \label{fig_hr_comparison}}
\end{figure}

In this study we do not attempt to analyse X-ray spectra of XMM0838+24 and XMM1226+12 in detail because given their faint X-ray fluxes it would not provide any selectivity to discriminate conventional X-ray object types.
However we compare their X-ray colors from the {\it XMM-Newton} source catalog to a sample of known AGN and ESO 243$-$49 HLX$-$1. 
On Fig.~\ref{fig_hr_comparison} we overplotted two ``corner'' states of ESO 243$-$49 HLX$-$1 and our HLX candidates on hardness diagrams of the known AGN population from \citet{lin12}. 
Clearly, broadband properties of ESO 243$-$49 HLX$-$1 do not allow its unambigous separation from AGN. 
At the same time selected HLX candidates are situated aside from ESO 243$-$49 HLX$-$1 and most AGN due to systematically harder HR2 and HR3 colors.
This however does not favor any particular interpretation of the nature of the candidates.

\subsection{Base sample properties}

We compute the number density of the HLX candidates from our base sample.
From total 98 HLX candidates we expect to have at least 21 true HLX.
Given the searched volume and solid angle this gives us HLX number density to be 1 object per $\approx 2 \times 10^5$~Mpc$^3$. This is approximately a factor of several thousand less frequent than ULXs \citep{swartz11}.
We can normalize this density to galactic halos.
In total there are 33,879 SDSS galaxies falling in the 3XMM-DR5 footprint (total observed part of the sky) which we included in our off-nuclear cross-match procedure.
Hence, the density of our HLX candidates is about $6 \times 10^{-4}$ per galaxy.
If they are all accreting IMBHs, this is too large a density for a population of ``naked'' IMBHs that captured stellar companions, estimated to be $\sim 10^{-7}$ per galaxy \citep[e.g.][]{kuranov07}.
At the same time this does not contradict the expected density of IMBHs that carry baryonic remnants remaining after inefficient galaxy mergers expected at the level of few $\times 10^{-2}$ per galaxy \citep{volonteri05}. 
Such light satellites are to have peripheral orbits almost unaffected by orbital decay, in line with significant galactocentric projected distances we observe in our candidates sample (15~kpc in the case of the source XMM0838+24 and 33~kpc for the source XMM1226+12).
We also note that like lower luminosity ULXs, two presented HLX candidates reside in late type galaxies, which is different from the ESO 243$-$49 HLX$-$1 host.
It is however too early to speculate on the environmental properties on the HLX candidates before the confirmation of their association with the assumed host galaxies.

\section{Conclusions}
\label{sec:conclusion}

In this study we presented a method to find off-nuclear X-ray sources in the vicinities of galaxies with known distances. 
Under the assumption that the X-ray sources are associated with the possible host galaxies, we searched for X-ray sources in the luminosity range typical for hyperluminous X-ray objects.
Constructed base sample of 98 X-ray sources has at most $71 \pm 11$ contaminants thus providing support for the existence of the HLX population.
We used public multi-wavelength surveys and data archives accessible through the Virtual Observatory to constrain spectral energy distributions of X-ray sources in a way that minimizes their chance to be conventional X-ray objects such as background AGN, blazars, star-forming galaxies or foreground Galactic objects.
In this paper we provide details on 2 HLX candidates from the base sample that satisfy criteria $f_{\rm X}/f_{\rm opt} > 100$ or $f_{\rm X}/f_{\rm NIR} > 40$ with existing data and discuss their possible nature.
Further follow up observations of the candidates are required to confirm them as HLXs, namely deeper and higher resolution X-ray imaging, optical identification and then optical spectroscopy in order to secure association with a host galaxy.

\acknowledgments

This research has made use of the VizieR catalogue access tool, CDS, Strasbourg, France.

This work is based in part on data obtained as part of the UKIRT Infrared Deep Sky Survey.

Funding for SDSS-III has been provided by the Alfred P. Sloan Foundation, the Participating Institutions, the National Science Foundation, and the U.S. Department of Energy Office of Science. The SDSS-III web site is http://www.sdss3.org/.

SDSS-III is managed by the Astrophysical Research Consortium for the Participating Institutions of the SDSS-III Collaboration including the University of Arizona, the Brazilian Participation Group, Brookhaven National Laboratory, Carnegie Mellon University, University of Florida, the French Participation Group, the German Participation Group, Harvard University, the Instituto de Astrofisica de Canarias, the Michigan State/Notre Dame/JINA Participation Group, Johns Hopkins University, Lawrence Berkeley National Laboratory, Max Planck Institute for Astrophysics, Max Planck Institute for Extraterrestrial Physics, New Mexico State University, New York University, Ohio State University, Pennsylvania State University, University of Portsmouth, Princeton University, the Spanish Participation Group, University of Tokyo, University of Utah, Vanderbilt University, University of Virginia, University of Washington, and Yale University.

IZ acknowledges support from RSCF grant no. 14-12-00146 for HLX sample selection and the grants MD-7355.2015.2 and RFBR 15-32-21062 for the archival data analysis.



{\it Facilities:} \facility{XMM-Newton}, \facility{Chandra}, \facility{Swift}, \facility{CDS VizieR}, \facility{CDS Simbad}, \facility{CDS Aladin}, \facility{TOPCAT}.

\appendix

\section{Details on the sample selection}
\label{sec:appendix_details}

As an effort to make our research results reproducible we give full technical details of the queries written in the Structured Query Language ({\sc sql}) we performed to construct the resulting sample of HLX candidates. These queries were launched to spectroscopical SDSS DR12 database in the SDSS {\it CasJobs}\footnote{\url{http://skyserver.sdss3.org/casjobs/}} interface. Column names are derived either from 3XMM-DR5 catalog or from SDSS DR12 catalog. Table names are self-explanatory.

After preparing the input X-ray source list (table {\tt mydb..xmm}, 309\,327 records) we first got the nearest SDSS spectral object within 2\arcmin\, from the X-ray coordinates of each X-ray object (82\,710 matches):

\begin{verbatim}
SELECT
  *, dbo.fGetNearestSpecObjIDEq(SC_RA, SC_DEC, 2.0) AS specobjid
INTO
  mydb..xmm_match_specdr12
FROM
  mydb..xmm
WHERE dbo.fGetNearestSpecObjIDEq(SC_RA, SC_DEC, 2.0) IS NOT NULL
\end{verbatim}

Then we used the following query to perform the off-nuclear cross-match (see Fig.~\ref{fig_xmatch} for the illustration of its positional criterion) between input X-ray list and list of SDSS spectral objects selected at previous step, to get additional photometrical information, calculate distance (discarding those further than 800~Mpc) and X-ray luminosity, filter only those X-ray objects with luminosities between $10^{41}$ and $10^44$\,\ergs, and not closer {\tt 3 * SC\_POSERR} arcsec (but at least 5\arcsec) to galaxy center and not further than {\tt 3 * SC\_POSERR + 2 * petroRad\_r} arcsec (373 matches):

\begin{verbatim}
SELECT 
  p.objid, p.ra AS sdss_ra, p.dec AS sdss_dec, p.petroRad_r, 
  p.petroR50_r, p.petroR90_r, x.*, 
  dbo.fDistanceEq(x.SC_RA, x.SC_DEC, p.ra, p.dec) * 60. AS separation,
  dbo.fCosmoDl(s.z,DEFAULT,DEFAULT,DEFAULT,DEFAULT,DEFAULT) AS distance, 
  x.SC_EP_8_FLUX * 4 * 3.1415926 * 
    power(dbo.fCosmoDl(s.z,DEFAULT,DEFAULT,DEFAULT,DEFAULT,DEFAULT) * 3.08e24, 2) AS luminosity
INTO 
  mydb..xmm_match_specdr12_offnuclear
FROM 
  photoObjAll AS p 
JOIN
  specObjAll AS s ON s.bestobjid = p.objid
JOIN 
  mydb..xmm_match_specdr12 AS x ON s.specobjid = x.specobjid
WHERE
  dbo.fDistanceEq(x.SC_RA, x.SC_DEC, p.ra, p.dec) * 60. > 3 * x.SC_POSERR
  AND dbo.fDistanceEq(x.SC_RA, x.SC_DEC, p.ra, p.dec) * 60. > 5.
  AND dbo.fDistanceEq(x.SC_RA, x.SC_DEC, p.ra, p.dec) * 60. < 3 * SC_POSERR + 2 * p.petrorad_r
  AND s.z > 0
  AND x.SC_EP_8_FLUX * 4 * 3.1415926 * 
    power(dbo.fCosmoDl(s.z,DEFAULT,DEFAULT,DEFAULT,DEFAULT,DEFAULT) * 3.08e24, 2) > 1e41
  AND x.SC_EP_8_FLUX * 4 * 3.1415926 * 
    power(dbo.fCosmoDl(s.z,DEFAULT,DEFAULT,DEFAULT,DEFAULT,DEFAULT) * 3.08e24, 2) < 1e44
  AND dbo.fCosmoDl(s.z,DEFAULT,DEFAULT,DEFAULT,DEFAULT,DEFAULT) < 800
\end{verbatim}

Adding additional constraint on X-ray flux: {\tt SC\_EP\_8\_FLUX > 5e-14} to this query make it to be our base sample of 98 HLX candidates.

As a last step we select high X-ray-to-optical ratio candidates. That is, we discard those sources that have optical counterparts in SDSS DR12 within {\tt 3 * SC\_POSERR}, other than the host spectroscopical galaxy itself:

\begin{verbatim}
SELECT
  *
INTO
  mydb..xmm_match_specdr12_offnuclear_fxfopt
FROM
  mydb..xmm_match_specdr12_offnuclear
WHERE
  SRCID NOT IN (
    SELECT
      x.SRCID
    FROM
      mydb..xmm_match_specdr12_offnuclear AS x, photoPrimary AS p
    WHERE
      p.objid = dbo.fGetNearestObjIDEq(x.SC_RA, x.SC_DEC, 3 * SC_POSERR / 60.)
      AND p.objid <> x.objid
    )
\end{verbatim}

This resulted in a list of 61 sources which we discarded hence keeping only 37 ``optically quiet'' sources for further more detailed inspection in this paper.

\bibliographystyle{apj}
\bibliography{hlx_search}

\clearpage

\end{document}